\documentclass[11pt]{article}

\usepackage[T1]{fontenc}
\usepackage[utf8]{inputenc}
\usepackage{lmodern}
\usepackage[margin=1in]{geometry}
\usepackage{amsmath}
\usepackage{amssymb}
\usepackage{amsfonts}
\usepackage{amsthm}
\usepackage[round,authoryear]{natbib}
\usepackage{graphicx}
\usepackage{xcolor}
\usepackage{booktabs}
\usepackage{enumitem}
\usepackage{caption}
\usepackage{microtype}

\usepackage[american]{babel}

\theoremstyle{plain}
\newtheorem{theorem}{Theorem}
\newtheorem{proposition}{Proposition}
\newtheorem{lemma}{Lemma}

\theoremstyle{definition}
\newtheorem{definition}{Definition}
\newtheorem{remark}{Remark}

\let\oldemph\emph
\renewcommand{\emph}[1]{\textcolor{blue!60!black}{\oldemph{#1}}}

\usepackage{hyperref}
\hypersetup{
        pdfencoding=auto,
        psdextra,
        colorlinks=true,
        citecolor=green!40!black,
        linkcolor=red!50!black,
        urlcolor=blue!80!black
    }
\DeclareMathOperator{\Ein}{Ein}

\title{How Should Candidates Be Weighted under Open-List Proportional Representation? Harmonic Weights and Dirichlet Order Statistics}
\author{Dariusz Stolicki\thanks{Center for Quantitative Political Science, Jagiellonian University, Krak\'ow, Poland. Corresponding author: \texttt{dariusz.stolicki@uj.edu.pl}.} \and Daria Boratyn\thanks{Center for Quantitative Political Science, Jagiellonian University, Krak\'ow, Poland.}}
\date{November 2026}

\begin{document}
\maketitle

\begin{abstract}
\noindent Candidates in open-list proportional representation systems are not equally prominent, and measurements that involve aggregation over candidates therefore require some form of weighting. We address the question of how those weights should be chosen. Since weights based on realized preference votes are unavailable in advance and endogenous to the electoral outcome, we seek a priori weights that depend on list position alone, and derive them as expected ordered vote shares under a probabilistic model of the division of the list vote among candidates. The natural starting point -- the uniform distribution on the simplex -- is unrealistic, as it prescribes the same degree of concentration of the vote for every list, and under it the leading candidates become more alike as lists lengthen. The other members of the symmetric Dirichlet family require a concentration parameter, which is unknown and which there is no reason to expect to be constant across countries, parties, and elections. Accordingly, we treat that parameter as random. We prove that under a scale-invariant prior restricted to lists with at least two effective candidates, the expected share of the $k$-th candidate decays as $1/k$, and we show that over the leading positions of a long list, up to one hundred of them, the resulting weights are within 1.4 to 3.4 percent, in total variation, of the harmonic weights $(k H_n)^{-1}$, where $H_n$ is the $n$-th harmonic number. No fixed value of the parameter yields similar weights. Moreover, under the uniform distribution itself, harmonic weights are exactly the normalized expected margins between consecutive candidates.


\end{abstract}

\section{\label{secIntro}Introduction}

\subsection{\label{secMotivation}Motivation}

Not all candidates in open-list proportional representation (OLPR) systems are equally prominent. Under OLPR, the number of seats won by a party is determined by the pooled vote of its list, while the candidates who fill those seats are determined by the preference votes cast for individuals \citep{CareyShugart95,ShugartEtAl05}. Those votes are divided very unevenly -- a few candidates attract most of the vote of their list, and many attract hardly any. In consequence, measurements that involve aggregation over candidates require some form of weighting, as an unweighted count treats the leader of a list and the candidate in its last position alike. In this paper we address the question of how those weights should be chosen.

Such measurements are common. Studies of candidate selection and of descriptive representation ask not only how many women or newcomers a party nominates, but also how good the positions are in which it places them \citep{HazanRahat06,RahatEtAl08,EsteveVolartBagues12,Luhiste15}. Candidate-based measures of party change and of candidate turnover follow the persons who stand for a party from one election to the next, and must decide how much the return of each of them counts \citep{KreuzerPettai03,SikkKoker19,KosowskaGastolSobolewskaMyslik23,MuytersMaddens23,SikkKoker23}. Moreover, the same weights serve wherever a value has to be attached to a list position, viz., in the measures of advancement on the list and of electoral vulnerability on which studies of political careers and of legislators' behavior rely \citep{AndreEtAl15,FivaRohr18}, in formal models of nomination, in which parties assign candidates of different quality to positions of different value \citep{BuisseretEtAl22,CrutzenEtAl24}, and in simulations of the allocation of seats within lists, which require a distribution of the list vote among candidates.

The unequal prominence of candidates has been handled in several ways, reviewed in Section~\ref{secEarlier}. These include cut-offs that separate viable positions from the rest, weights that decline affinely with rank, estimated probabilities of election, and smooth functions of relative list placement such as the logistic weights of \citet[ch.~3]{SikkKoker23}. Most of these schemes draw on the result of the party at the previous or the current election. An exception is the reciprocal of the list position, proposed as a weight by \citet{FlisEtAl22} and adopted by \citet{FlisLorenc24}, which coincides up to normalization with the harmonic weights studied in this paper, but has been used as an approximation, without derivation. Weights based on realized preference votes capture the heterogeneity directly. Nevertheless, they are not available before the election or for lists that did not compete under open-list rules, and they are endogenous whenever the quantity of interest is itself related to the electoral outcome.

We therefore seek weights that can be assigned a priori, as a function of list position alone. List position is informative for two reasons. First, parties tend to place their strongest candidates at the top of the list. Second, candidates placed higher on the ballot receive more preference votes by virtue of their position \citep{FaasSchoen06,Lutz10,Marcinkiewicz14}. We take the weight of a position to be the share of the list vote that its occupant can be expected to receive, and we derive it from a probabilistic model of the division of that vote, assuming that the order of the list reflects the order of the candidates' votes.

\subsection{\label{secContribution}Contribution}

\textbf{First.} We show that the natural starting point for such a model -- the uniform distribution on the probability simplex, under which all divisions of the list vote among the candidates are equally likely -- is unrealistic. That distribution is the counterpart of the impartial anonymous culture of social choice theory \citep{GehrleinFishburn76}, and the expected ordered coordinates of a uniformly distributed point of the simplex have been used to model the weights of players in weighted voting games \citep{BoratynEtAl20}. Nevertheless, it prescribes the same degree of concentration for every list, with an effective number of candidates equal to about half of the length of the list; under it the leading candidates become more alike as lists lengthen; and, judging by the dispersion of candidates' votes reported by \citet{FortunatoCastellano07}, votes are in fact divided more unequally than it allows.

\textbf{Second.} We show that those shortcomings are not remedied by any other member of the symmetric Dirichlet family, which arises as the large-electorate limit of the P\'olya--Eggenberger urn model \citep{Berg85}, as long as its parameter is fixed. With a fixed parameter per candidate, the profile of expected shares is too flat among the leading candidates and too steep among the trailing ones (Proposition~\ref{prop:shape}), and the leading candidates still become more alike as the list grows. Moreover, the concentration parameter is unknown, and there is no reason to expect it to be constant across countries, parties, and elections, as the extent to which votes are concentrated on a few candidates can be expected to depend on the electoral rules, which shape the incentives to cultivate a personal vote \citep{CareyShugart95,ShugartEtAl05}, on the party, and on the candidates that it fields. Any fixed value is therefore a substantive assumption of the kind that an a priori scheme is meant to avoid.

\textbf{Third.} We treat the concentration parameter as random. To this end, we express it as the total concentration of the list, i.e., the sum of the Dirichlet parameters of its candidates, and we assume that its distribution does not depend on the length of the list. As the list grows with the total concentration held fixed, the ordered shares converge to the Poisson--Dirichlet distribution \citep{Kingman75}, which has been studied extensively in combinatorics and population genetics \citep{SheppLloyd66,Watterson76,Griffiths88,PitmanYor97}, and the ratios between the expected shares of the leading candidates no longer tend to one (Proposition~\ref{prop:sparse}). We prove that if the total concentration is assigned the scale-invariant prior, which expresses ignorance of the order of magnitude of a positive quantity \citep{Jeffreys46,Jaynes68}, those shares decay as $1/k$ (Theorem~\ref{thm:random}). From the second position on, the share of the $k$-th candidate is proportional to $(k-1)^{-1}$ up to a geometrically small correction, while the share of the leader is finite and is determined by the lower bound of the prior, which we set so as to admit all lists with at least two effective candidates. The weights obtained in this way are closely approximated by the \emph{harmonic weights}, under which the $k$-th of $n$ candidates receives weight $(k H_n)^{-1}$, where $H_n$ is the $n$-th harmonic number (Definition~\ref{def:harmonic}). Over any number of leading positions up to one hundred, the two sets of weights differ by 1.4 to 3.4 percent of the total weight, whereas no fixed value of the parameter leads to similar weights. Harmonic weights thus emerge as the a priori weights appropriate when neither the allocation of votes among candidates nor its degree of concentration is known.

\textbf{Fourth.} We show that under the uniform distribution itself the harmonic weights are exactly the normalized expected margins between consecutive candidates, and that within the symmetric Dirichlet family this property holds for the uniform distribution only (Proposition~\ref{prop:margins}).

The above derivation has limits, which are stated in Sections~\ref{secRandom} and~\ref{secDiscussion}. In particular, the weight of the leader relative to the other positions depends on the lower bound of the prior, the approximation applies to the leading positions of a long list, and on a list considered in isolation it requires a prior that extends to nearly even divisions of the vote.

The remainder of the paper is organized as follows. Section~\ref{secEarlier} reviews existing schemes for weighting candidates on ordered lists. Section~\ref{secModel} introduces the model and a representation of expected ordered shares on which the remainder of the paper relies. Section~\ref{secUniform} deals with the uniform distribution, Section~\ref{secFixed} with Dirichlet distributions with a fixed parameter, and Section~\ref{secRandom} with random concentration. Section~\ref{secDiscussion} returns to the choice of weights and discusses the limitations of the analysis. Numerical methods are described in the Appendix.

\subsection{\label{secNotation}Notation}

Throughout the paper, the symbol $:=$ denotes equality by definition, $\mathbb{N}_{+} := \{1, 2, \dots\}$, and for any $n \in \mathbb{N}_{+}$ we denote $[n] := \{1, \dots, n\}$. We use boldface for vectors and capital letters for random variables. For any $n \in \mathbb{N}_{+}$, let
\[
    \Delta_n := \left\{\mathbf{x} \in \mathbb{R}_{\ge 0}^{n} : \sum_{i=1}^{n} x_i = 1\right\}
\]
be the $(n-1)$-dimensional probability simplex. For any $\mathbf{x} \in \mathbb{R}^{n}$, we denote by $x_{(1)} \ge \dots \ge x_{(n)}$ its coordinates arranged in nonincreasing order, so that $x_{(k)}$ is the $k$-th largest coordinate of $\mathbf{x}$. Expectation and probability are denoted by $\mathbb{E}(\cdot)$ and $\Pr(\cdot)$, respectively. The relation $f(x) \sim g(x)$ as $x \rightarrow a$ signifies that $f(x)/g(x) \rightarrow 1$ as $x \rightarrow a$.

For any $n \in \mathbb{N}_{+}$, let $H_n := \sum_{j=1}^{n} 1/j$ be the $n$-th harmonic number, and let $H_0 := 0$. Moreover, let $\Gamma(a, x) := \int_{x}^{\infty} t^{a-1} e^{-t} \,dt$, where $a > 0$ and $x \ge 0$, be the upper incomplete gamma function, so that $\Gamma(a) := \Gamma(a, 0)$, let $E_{1}(x) := \int_{x}^{\infty} t^{-1} e^{-t} \,dt$, where $x > 0$, be the exponential integral, and let $\gamma$ be the Euler--Mascheroni constant.

\section{\label{secEarlier}Existing Weights for Candidates on Ordered Lists}

Schemes that assign unequal weights to the candidates of an ordered list can be divided into five groups, according to how the value of a list position is defined.

\emph{Cut-offs.} The simplest schemes count only the candidates above a threshold, all of them equally. The threshold is either a fixed fraction of the list, such as its top quarter \citep{SikkKoker19} or its top tenth \citep[see][]{KosowskaGastolSobolewskaMyslik23}, or the number of positions regarded as safe \citep{HazanRahat06,FivaRohr18}, eligible \citep{PutMaddens13}, or viable \citep{Luhiste15}. In the first of these studies the number of safe positions is the number of seats that the party won at the previous election.

\emph{Weights declining affinely with rank.} \citet{HazanRahat06} refine the cut-off by valuing the safe positions unequally: the last safe position is worth one point and each higher position one point more. Their weighted index of representation is the corresponding weighted share of women among the safe positions, and a weighted index of competition, in which the threshold is the number of competing incumbents, is constructed in the same way for non-incumbents; both kinds of index are applied by \citet{RahatEtAl08}. \citet{FivaRohr18} standardize the position of the $k$-th of $n$ candidates as $1 - (k-1)/(n-1)$ over the whole list, and \citet{JankowskiRehmert25} score a list of six positions from six points for the first position down to one point for the last. \citet[ch.~3]{SikkKoker23} consider weights of this form and reject them as giving too much weight to candidates without prospects of election.

\emph{Probabilities of election.} A third group of schemes values a position by the probability that its occupant wins a seat, estimated from earlier elections. \citet{ManowNistor09} compute this probability for every position on the state lists of German parties; \citet{EsteveVolartBagues12} estimate it from the position and the previous support of the party; \citet{StoffelSieberer18} obtain re-election probabilities from multilevel logistic regressions; and the measure of electoral vulnerability of \citet{AndreEtAl15} serves a related purpose. The logistic weights of \citet[ch.~3]{SikkKoker23} are close in spirit. A smooth function of relative list placement with the midpoint set at the list's vote share, they decline mildly at the top of the list, fastest around the position at which the seats of the party can be expected to run out, and slowly at the bottom. \citet{MuytersMaddens23} likewise use a weighted measure of candidate turnover to study renewal in high-ranked positions.

\emph{Unspecified decreasing weights.} In formal models of nomination, list position may enter through weights whose form is left open. For instance, \citet{CrutzenEtAl24} let the electoral output of a party be a weighted sum of the efforts of its candidates, with weights that are weakly decreasing in list position.

\emph{Reciprocal weights.} \citet{FlisEtAl22} weight the position of a candidate on the party list by its reciprocal, i.e., by $1/k$ for the $k$-th position. \citet{FlisLorenc24} adopt that rule in their study of political career paths in four elections to the Polish Sejm, and compare the share of each group of candidates in the total weight so obtained with its share in the number of candidates. Up to normalization, these are the harmonic weights of Definition~\ref{def:harmonic}. The rule is presented as an approximation of the effect of list placement, and it is not derived.

\begin{remark} \label{rem:earlier}
    Note that, with the exception of fixed-fraction cut-offs, of affine weights defined over the whole list, and of reciprocal weights, the above schemes require information beyond the list itself, viz., the result of the party at the previous election or, in the case of the logistic weights, at the election in question. Moreover, nearly all of them value a position by the prospect of winning a seat. That is the natural criterion under closed lists, where the order of the list decides who is elected. Under open lists, where the seats of a list are allocated by preference votes and the order of the list is only a starting point, the natural counterpart is the share of the list vote that a candidate can be expected to receive, and none of the above schemes is derived from a model of how that vote is divided. Finally, the reciprocal weights of \citet{FlisEtAl22} and \citet{FlisLorenc24} are the only ones that coincide, up to normalization, with the harmonic weights studied in this paper, and they are adopted as an approximation rather than derived.
\end{remark}

\section{\label{secModel}Model}

Fix any $n \in \mathbb{N}_{+}$ such that $n \ge 2$, and consider a list of $n$ candidates. Let $\mathbf{V} := (V_1, \dots, V_n)$ be the random vector of the shares of the list vote obtained by those candidates. Clearly, $\mathbf{V}$ takes values in $\Delta_n$. Unless stated otherwise, we assume that:

\begin{enumerate}[topsep=0pt,leftmargin=*,label=(\textbf{A\arabic*})]
    \item \label{A1} (sorting) the candidates appear on the list in nonincreasing order of their vote shares, i.e., the share of the candidate in the $k$-th position is $V_{(k)}$ for every $k \in [n]$;

    \item \label{A2} (symmetric Dirichlet shares) the vector $\mathbf{V}$ has the symmetric Dirichlet distribution with parameter $\alpha > 0$, i.e., the distribution on $\Delta_n$ with density proportional to $\prod_{i=1}^{n} v_i^{\alpha - 1}$.
\end{enumerate}

\begin{remark} \label{rem:assumptions}
    Assumption \textbf{A1} provides the link between ordered shares and list positions. It is an idealization, to which we return in Section~\ref{secDiscussion}. Assumption \textbf{A2} is exchangeable, i.e., it carries no information about individual candidates, which is what an a priori scheme requires. For $\alpha = 1$ it yields the uniform distribution on $\Delta_n$. For $\alpha < 1$ the probability mass moves toward the vertices of the simplex, so that the vote is typically concentrated on few candidates, while for $\alpha > 1$ it moves toward the barycenter.
\end{remark}

\begin{definition}[Expected-share profile] \label{def:profile}
    Assume that \textbf{A2} holds. The \emph{expected-share profile} of a list of $n$ candidates is the vector $\boldsymbol{\pi}^{(n)}(\alpha) := \left(\pi^{(n)}_1(\alpha), \dots, \pi^{(n)}_n(\alpha)\right) \in \Delta_n$, where $\pi^{(n)}_k(\alpha) := \mathbb{E}(V_{(k)})$ for every $k \in [n]$.
\end{definition}

Note that under \textbf{A1} the $k$-th coordinate of the expected-share profile is the expected share of the candidate in the $k$-th position, i.e., the weight that an analyst who knows the model but not the result would assign to that position.

\begin{definition}[Total concentration] \label{def:concentration}
    Assume that \textbf{A2} holds. The \emph{total concentration} of a list of $n$ candidates is $\theta := n\alpha$, i.e., the sum of the Dirichlet parameters of its candidates.
\end{definition}

The parameter of the symmetric Dirichlet distribution governs the concentration of the vote, and it can be expressed either per candidate, as $\alpha$, or for the list as a whole, as $\theta$. Indeed, since $\mathbb{E}\left(\sum_{i=1}^{n} V_i^2\right) = (\alpha + 1)/(n\alpha + 1)$, the effective number of candidates in the sense of \citet{LaaksoTaagepera79}, evaluated at the expected sum of squared shares, equals
\begin{equation}
    N_{\mathrm{eff}} := \frac{n\alpha + 1}{\alpha + 1}.  \label{eq:effNumber}
\end{equation}
In particular, $N_{\mathrm{eff}} = (n+1)/2$ under the uniform distribution.

\begin{definition}[Harmonic weights] \label{def:harmonic}
    The \emph{harmonic weights} for a list of $n$ candidates are the coordinates of the vector $\mathbf{w}^{(n)} := \left(w^{(n)}_1, \dots, w^{(n)}_n\right) \in \Delta_n$, where
    \begin{equation}
        w^{(n)}_k := \frac{1}{k H_n} \qquad \text{for every } k \in [n].  \label{eq:harmonic}
    \end{equation}
\end{definition}

The harmonic scheme has no free parameters, and the relative weight of any two positions, $w^{(n)}_k / w^{(n)}_l = l/k$, does not depend on the length of the list. The weights themselves do depend on it, as they are normalized over the list, e.g., the weight of the leader equals $1/H_n$.

\begin{remark} \label{rem:dhondt}
    The sequence $1, 1/2, 1/3, \dots$ is also that of the quotients of the Jefferson--D'Hondt method. Indeed, on a closed list, the $k$-th candidate is elected if and only if the vote of the list divided by $k$ is among the $M$ largest quotients of all lists, where $M$ is the number of seats \citep{Pukelsheim17}.
\end{remark}

\begin{definition}[Total variation distance] \label{def:tv}
    Fix any $m \in \mathbb{N}_{+}$. The \emph{total variation distance} between two weight profiles $\mathbf{p}, \mathbf{q} \in \Delta_m$ is
    \[
        d(\mathbf{p}, \mathbf{q}) := \frac{1}{2} \sum_{k=1}^{m} |p_k - q_k|,
    \]
    i.e., the share of the total weight that would have to be moved to turn one profile into the other.
\end{definition}

The expected-share profile can be expressed through the order statistics of independent gamma variables. Let $G_1, \dots, G_n$ be independent random variables having the gamma distribution with shape $\alpha$ and unit scale. Let $f_\alpha$, $F_\alpha$, and $Q_\alpha$ be, respectively, the density, the distribution function, and the quantile function of that distribution, and let $h_\alpha := f_\alpha / (1 - F_\alpha)$ be its hazard rate. Recall that $G_{(1)} \ge \dots \ge G_{(n)}$ are the values of $G_1, \dots, G_n$ arranged in nonincreasing order, and let $G_{(n+1)} := 0$.

\begin{lemma}[Gamma representation] \label{lem:gamma}
    Assume that \textbf{A2} holds. Then for every $k \in [n]$,
    \begin{equation}
        \pi^{(n)}_k(\alpha) = \frac{\mathbb{E}(G_{(k)})}{n\alpha}  \label{eq:levels}
    \end{equation}
    and
    \begin{equation}
        \mathbb{E}(G_{(k)}) - \mathbb{E}(G_{(k+1)}) = \binom{n}{k} \int_{0}^{\infty} F_\alpha(x)^{n-k} \left(1 - F_\alpha(x)\right)^{k} \,dx = \frac{1}{k} \mathbb{E}\left(\frac{1}{h_\alpha(G_{(k)})}\right).  \label{eq:spacings}
    \end{equation}
\end{lemma}

\begin{proof}
    Fix any $k \in [n]$. Recall that $\mathbf{V}$ has the distribution of $(G_1/S, \dots, G_n/S)$, where $S := \sum_{i=1}^{n} G_i$, and that $S$ is independent of that vector \citep[ch.~49]{KotzEtAl00}. Hence $\mathbb{E}(G_{(k)}) = \mathbb{E}(V_{(k)}) \mathbb{E}(S) = n\alpha \pi^{(n)}_k(\alpha)$, as desired. Note next that
    \[
        \mathbb{E}(G_{(k)}) - \mathbb{E}(G_{(k+1)}) = \int_{0}^{\infty} \left(\Pr(G_{(k)} > x) - \Pr(G_{(k+1)} > x)\right) \,dx,
    \]
    and that the integrand is the probability that exactly $k$ of the $n$ variables exceed $x$, which yields the first equality in (\ref{eq:spacings}) \citep[cf.][ch.~3]{DavidNagaraja03}. The substitution $u := 1 - F_\alpha(x)$, for which $dx = -(u h_\alpha(x))^{-1} \,du$, turns the integral into
    \[
        \int_{0}^{1} \binom{n}{k} u^{k-1} (1 - u)^{n-k} \frac{1}{h_\alpha(Q_\alpha(1 - u))} \,du.
    \]
    Since the function $u \mapsto \binom{n}{k} u^{k-1} (1-u)^{n-k}$ is $1/k$ times the density of the beta distribution with parameters $k$ and $n - k + 1$, i.e., of the distribution of $1 - F_\alpha(G_{(k)})$, the second equality follows.
\end{proof}

All Dirichlet expected-share profiles reported below were computed from (\ref{eq:levels}) by one-dimensional quadrature, without simulation (see the Appendix for details).

\section{\label{secUniform}The Uniform Distribution}

In the absence of any information about the candidates of a list, the natural starting point is the uniform distribution on $\Delta_n$, i.e., the case $\alpha = 1$ of \textbf{A2}, under which all divisions of the list vote are equally likely. Its expected-share profile follows from the expected margins between consecutive candidates. Let $\pi^{(n)}_{n+1}(\alpha) := 0$. By Lemma~\ref{lem:gamma}, for every $\alpha > 0$ and every $k \in [n]$ the expected margin by which the $k$-th candidate leads the next one equals
\begin{equation}
    \pi^{(n)}_k(\alpha) - \pi^{(n)}_{k+1}(\alpha) = \frac{\rho^{(n)}_k(\alpha)}{k n \alpha}, \qquad \text{where} \qquad \rho^{(n)}_k(\alpha) := \mathbb{E}\left(\frac{1}{h_\alpha(G_{(k)})}\right).  \label{eq:margin}
\end{equation}
Thus, the expected margins are harmonic up to the factors $\rho^{(n)}_k(\alpha)$, i.e., up to the expected reciprocal hazard rates at the ordered observations.

\begin{proposition}[Expected margins] \label{prop:margins}
    Assume that \textbf{A2} holds. Then:
    \begin{enumerate}
        \item[(i)] if $\alpha = 1$, then $\rho^{(n)}_k(1) = 1$ for every $k \in [n]$, and in consequence
        \begin{equation}
            \pi^{(n)}_k(1) = \frac{1}{n} \sum_{j=k}^{n} \frac{1}{j} \qquad \text{and} \qquad w^{(n)}_k = \frac{\pi^{(n)}_k(1) - \pi^{(n)}_{k+1}(1)}{\pi^{(n)}_1(1)} \qquad \text{for every } k \in [n];  \label{eq:uniform}
        \end{equation}

        \item[(ii)] if $\alpha < 1$, then $\rho^{(n)}_k(\alpha) < 1$ and $\rho^{(n)}_k(\alpha)$ is strictly decreasing in $k$, whereas if $\alpha > 1$, then $\rho^{(n)}_k(\alpha) > 1$ and $\rho^{(n)}_k(\alpha)$ is strictly increasing in $k$; in either case the normalized expected margins are not harmonic;

        \item[(iii)] for every $\alpha > 0$ and every fixed $k \in \mathbb{N}_{+}$, $\rho^{(n)}_k(\alpha) \rightarrow 1$ as $n \rightarrow \infty$.
    \end{enumerate}
\end{proposition}

\begin{proof}
    (i) Note that the hazard rate of the exponential distribution is $h_1 \equiv 1$. Accordingly, the first formula in (\ref{eq:uniform}) follows by summing (\ref{eq:margin}) over the positions from the $k$-th to the last, and the second one from $\pi^{(n)}_1(1) = H_n / n$.

    (ii) Recall that the hazard rate of the gamma distribution is strictly decreasing for $\alpha < 1$ and strictly increasing for $\alpha > 1$, in both cases with limit one at infinity \citep{BarlowProschan75}. Fix any $\alpha < 1$. Then the function $1/h_\alpha$ is strictly increasing and smaller than one, and since $G_{(k+1)} < G_{(k)}$ almost surely, we obtain $\rho^{(n)}_{k+1}(\alpha) < \rho^{(n)}_k(\alpha) < 1$. The case $\alpha > 1$ is symmetric. Finally, since $\pi^{(n)}_1(\alpha) = \sum_{j=1}^{n} (j n \alpha)^{-1} \rho^{(n)}_j(\alpha)$, the normalized margins are harmonic only if $\rho^{(n)}_k(\alpha)$ does not depend on $k$.

    (iii) For $\alpha = 1$ the claim follows from (i). Fix any $\alpha \neq 1$ and any $k \in \mathbb{N}_{+}$. As $n$ grows, $G_{(k)}$ is stochastically nondecreasing and tends to infinity in probability, while $1/h_\alpha(x) \rightarrow 1$ as $x \rightarrow \infty$. If $\alpha < 1$, the claim follows by bounded convergence. If $\alpha > 1$, the function $1/h_\alpha$ is decreasing, so that $1/h_\alpha(G_{(k)})$ is nonincreasing in $n$ when the samples are nested. Since it is integrable for $n = k$ by (\ref{eq:spacings}), the claim follows by monotone convergence.
\end{proof}

By Proposition~\ref{prop:margins}(i), the harmonic weights are already present in the uniform model, although not as expected shares. Indeed, the harmonic weight of the $k$-th position is the expected lead of its occupant over the next candidate, expressed as a fraction of the leader's expected share.

\begin{remark} \label{rem:classical}
    The first formula in (\ref{eq:uniform}) is classical \citep[see][for the use of the expected shares under the uniform distribution as average voting weights]{BoratynEtAl20}. Note that those shares are not harmonic.
\end{remark}

\begin{remark} \label{rem:marginScope}
    By Proposition~\ref{prop:margins}(ii), the above identity is specific to the uniform distribution, and the departures from it are substantial. For instance, for $n = 10$, $\rho^{(n)}_k(1/2)$ falls from $0.81$ at the top of the list to $0.12$ at the bottom, $\rho^{(n)}_k(2)$ rises from $1.24$ to $4.66$, and in both cases the normalized expected margins are at a total variation distance of $0.12$ from the harmonic weights. Nevertheless, by Proposition~\ref{prop:margins}(iii), the margins between the leading candidates of a long list are approximately harmonic whatever the value of $\alpha$, although the convergence is slow, e.g., $\rho^{(n)}_1(1/2)$ equals $0.81$, $0.87$, and $0.90$ for $n = 10$, $30$, and $100$, respectively.
\end{remark}

Nevertheless, the uniform distribution is not a realistic model of the intra-list vote. First, it prescribes one degree of concentration for all lists. Indeed, by (\ref{eq:effNumber}), the effective number of candidates equals $(n+1)/2$, i.e., about half of the length of the list, whatever the country, the party, or the election. Second, under it the leading candidates become more alike as the list lengthens. Indeed, the expected share of the second candidate relative to that of the leader, $1 - 1/H_n$, rises from $0.66$ on a list of ten candidates to $0.81$ on a list of one hundred, while the leader's expected share, $H_n/n$, falls from $0.293$ to $0.052$, i.e., by a factor of $5.6$. Yet a list can be lengthened by candidates who are not expected to attract many votes, and there is no reason why their presence should alter the standing of the leaders relative to one another. Third, votes are in fact divided more unequally than the uniform distribution allows. Under that distribution, the variance of a candidate's share relative to the average share on the list equals $(n-1)/(n+1)$, which is less than one, whereas the lognormal distribution that \citet{FortunatoCastellano07} fit to this quantity for elections in Italy, Poland, and Finland has a variance of about $1.9$.

\section{\label{secFixed}Dirichlet Distributions with a Fixed Parameter}

The uniform distribution is the member $\alpha = 1$ of the symmetric Dirichlet family, and the obvious remedy for its excessive evenness is a smaller parameter. A value of $\alpha$ then has to be chosen, and the choice matters. For instance, on a list of thirty candidates the leader's expected share is $0.27$, $0.19$, $0.13$, and $0.10$ for $\alpha = 1/4$, $1/2$, $1$, and $2$, respectively. Moreover, two properties of the expected-share profile hold for every fixed value of $\alpha$. The first one concerns the shape of the profile, which is best described by the proportional rate at which it declines along the list.

\begin{definition}[Elasticity] \label{def:elasticity}
    Fix any $m \in \mathbb{N}_{+}$, let $\mathbf{p} \in \Delta_m$ be a weight profile with positive coordinates, and let $p_{m+1} := 0$. The \emph{elasticity} of $\mathbf{p}$ at the $k$-th position, where $k \in [m]$, is
    \[
        \varepsilon_k(\mathbf{p}) := \frac{k (p_k - p_{k+1})}{p_k}.
    \]
\end{definition}

For the harmonic weights we have $\varepsilon_k\left(\mathbf{w}^{(n)}\right) = k/(k+1)$ for every $k < n$, which is close to one except at the very top of the list. More generally, for a sequence proportional to $k^{-s}$, where $s > 0$, the analogous quantity tends to $s$ as $k \rightarrow \infty$. By Lemma~\ref{lem:gamma}, the elasticity of the expected-share profile equals
\begin{equation}
    \varepsilon_k\left(\boldsymbol{\pi}^{(n)}(\alpha)\right) = \frac{\mathbb{E}\left(1/h_\alpha(G_{(k)})\right)}{\mathbb{E}(G_{(k)})}.  \label{eq:elasticity}
\end{equation}
Recall that for large $n$ the $k$-th largest observation concentrates around $x_k := Q_\alpha(1 - k/(n+1))$ \citep[ch.~4]{DavidNagaraja03}. Accordingly, the right-hand side of (\ref{eq:elasticity}) is then approximately $(x_k h_\alpha(x_k))^{-1}$.

\begin{proposition}[Shape of the profile] \label{prop:shape}
    For every $\alpha > 0$ the function $x \mapsto x h_\alpha(x)$ is strictly increasing on $(0, \infty)$, with limits $0$ at zero and $\infty$ at infinity.
\end{proposition}

\begin{proof}
    Fix any $\alpha > 0$. Since $(\log (1 - F_\alpha))' = -h_\alpha$, the derivative of $\log (x h_\alpha(x))$ equals $\alpha/x - 1 + h_\alpha(x)$. If $\alpha \le 1$, the hazard rate is nonincreasing with limit one, wherefore $h_\alpha \ge 1$ and the derivative is at least $\alpha/x$. If $\alpha > 1$, then integrating by parts we obtain $\Gamma(\alpha, x) = x^{\alpha-1} e^{-x} + (\alpha - 1) \Gamma(\alpha - 1, x)$. Since $\Gamma(\alpha - 1, x) \le \Gamma(\alpha, x)/x$, it follows that
    \[
        h_\alpha(x) = \frac{x^{\alpha-1} e^{-x}}{\Gamma(\alpha, x)} \ge 1 - \frac{\alpha - 1}{x} > 1 - \frac{\alpha}{x}.
    \]
    Thus, the derivative is positive in both cases. Finally, the limits follow from $h_\alpha(x) \sim x^{\alpha-1}/\Gamma(\alpha)$ as $x \rightarrow 0$ and $h_\alpha(x) \rightarrow 1$ as $x \rightarrow \infty$.
\end{proof}

\begin{figure}[tb]
    \centering
    \includegraphics[width=0.84\textwidth]{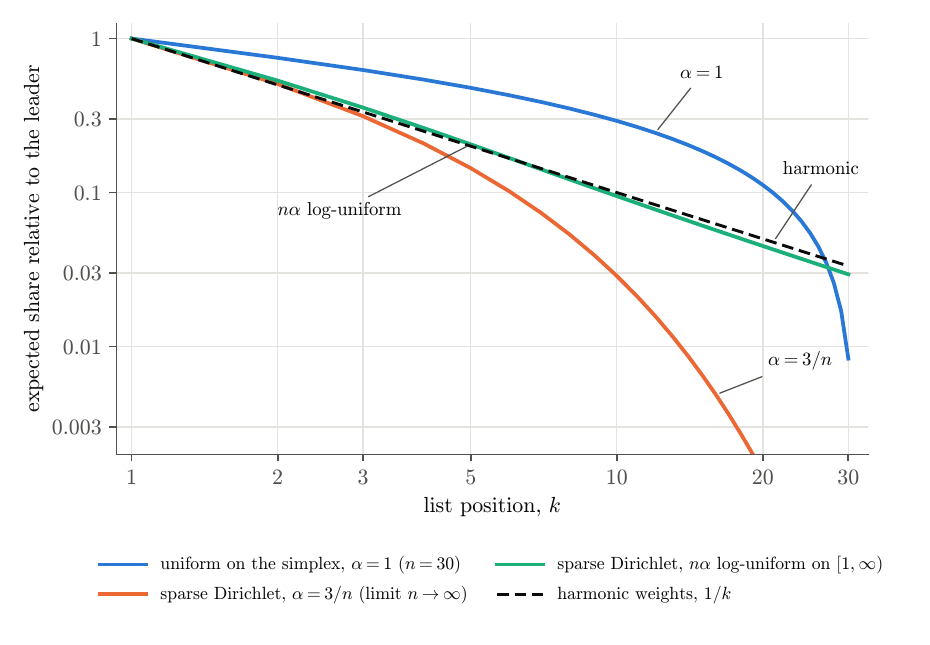}
    \caption{Expected share of the $k$-th candidate relative to the leader under the uniform distribution on the simplex ($\alpha = 1$, $n = 30$), in the sparse limit with total concentration $n\alpha = 3$, and in the sparse limit with $n\alpha$ log-uniform on $[1, \infty)$, compared with the harmonic weights. Both axes are logarithmic.}
    \label{fig:profiles}
\end{figure}

It follows from Proposition~\ref{prop:shape} that, in the large-$n$ approximation, the elasticity of a Dirichlet expected-share profile increases strictly from zero at the top of the list to infinity at the bottom. Thus, whatever the value of $\alpha$, such a profile is nearly flat among the leading candidates and collapses among the trailing ones. On doubly logarithmic axes it is concave (Figure~\ref{fig:profiles}, blue curve), whereas a profile that declines as a power of the position is a straight line. The elasticity equals one at a single point, located at the fraction $u^*_\alpha := 1 - F_\alpha(x^*)$ of the list, where $x^*$ is the solution of $x^* h_\alpha(x^*) = 1$. This fraction equals $0.14$, $0.23$, $1/e \approx 0.37$, and $0.52$ for $\alpha = 1/4$, $1/2$, $1$, and $2$, respectively. In consequence, lowering $\alpha$ steepens the profile at the leading positions only at the cost of a faster collapse at the lower ones.

The second property concerns list length. By Proposition~\ref{prop:margins}(iii) and (\ref{eq:margin}), for fixed $\alpha$ and $k$ the difference $\mathbb{E}(G_{(1)}) - \mathbb{E}(G_{(k)})$ converges to $H_{k-1}$ as $n \rightarrow \infty$, whereas $\mathbb{E}(G_{(1)})$ grows without bound, at the rate $\log n$ \citep[ch.~3]{EmbrechtsEtAl97}. Hence, the leader's expected share vanishes at the rate $(n \alpha)^{-1} \log n$, and the ratios between the expected shares of the leading candidates tend to one as the list grows. Thus, the property noted for the uniform distribution in Section~\ref{secUniform} is shared by every member of the family with a fixed $\alpha$: as the list lengthens, its leading candidates become more alike.

\section{\label{secRandom}Random Concentration}

A value of the parameter thus has to be chosen. However, an analyst who assigns weights a priori knows the concentration of the intra-list vote no better than its allocation among candidates. Moreover, the parameter is not a constant that could be determined once and for all, as there is no reason to expect it to be the same across countries, parties, and elections. Accordingly, we let it be random.

We express the parameter as the total concentration $\theta$ (Definition~\ref{def:concentration}) rather than as $\alpha$, and we assume that the distribution of $\theta$ does not depend on the length of the list. Note that if $\theta$ rather than $\alpha$ is held fixed as the list grows, i.e., if $\alpha = \theta/n$, then by (\ref{eq:effNumber}) the effective number of candidates tends to $1 + \theta$, and, as shown below, the ratios between the expected shares of the leading candidates no longer tend to one. We refer to this limit as the \emph{sparse regime}, and we begin by describing the expected shares in it.

\begin{definition}[Sparse profile] \label{def:sparse}
    Fix any $\theta > 0$. The \emph{sparse profile} with total concentration $\theta$ is the sequence $\boldsymbol{\ell}(\theta) := (\ell_k(\theta))_{k \in \mathbb{N}_{+}}$, where
    \begin{equation}
        \ell_k(\theta) := \frac{\theta^{k-1}}{(k-1)!} \int_{0}^{\infty} E_{1}(x)^{k-1} e^{-\theta E_{1}(x) - x} \,dx.  \label{eq:sparseProfile}
    \end{equation}
\end{definition}

\begin{proposition}[Sparse regime] \label{prop:sparse}
    Fix any $\theta > 0$, and assume that \textbf{A2} holds with $\alpha = \theta/n$. Then for every $k \in \mathbb{N}_{+}$,
    \begin{equation}
        \lim_{n \rightarrow \infty} \pi^{(n)}_k(\theta/n) = \ell_k(\theta).  \label{eq:sparseLimit}
    \end{equation}
    Moreover, $\sum_{k=1}^{\infty} \ell_k(\theta) = 1$, and for every $k \in \mathbb{N}_{+}$,
    \begin{equation}
        \ell_k(\theta) \ge \frac{e^{-\gamma}}{\theta} \left(\frac{\theta}{1 + \theta}\right)^{k},  \label{eq:tail}
    \end{equation}
    with the ratio of the two sides tending to one as $k \rightarrow \infty$.
\end{proposition}

\begin{proof}
    As $n \rightarrow \infty$ with $\alpha = \theta/n$, the ordered shares converge in distribution to the Poisson--Dirichlet distribution with parameter $\theta$ \citep{Kingman75}, and since they are bounded, their expectations converge as well. Recall that the limit is the distribution of the sequence $(J_1/S, J_2/S, \dots)$, where $J_1 \ge J_2 \ge \dots$ are the ranked points of a Poisson process on $(0, \infty)$ with intensity $(\theta/x) e^{-x} \,dx$ and $S$ is their sum. Moreover, $S$ has the gamma distribution with shape $\theta$ and is independent of the normalized sequence. Hence, the limit in (\ref{eq:sparseLimit}) equals $\mathbb{E}(J_k)/\theta$. The number of points exceeding $x$ has the Poisson distribution with expectation $\theta E_{1}(x)$, so that $J_k$ has density $((k-1)!\,x)^{-1} \theta^{k} E_{1}(x)^{k-1} e^{-\theta E_{1}(x) - x}$, which yields (\ref{eq:sparseLimit}). Summing over $k$ under the integral, we obtain $\sum_{k=1}^{\infty} \ell_k(\theta) = \int_{0}^{\infty} e^{-x} \,dx = 1$.

    To prove (\ref{eq:tail}), let $T_k$ be the sum of $k$ independent unit exponential variables. By the same Poisson property, $J_k$ has the distribution of $E_{1}^{-1}(T_k/\theta)$. Recall that $E_{1}(x) = -\gamma - \log x + \Ein(x)$, where $\Ein(x) := \int_{0}^{x} t^{-1} (1 - e^{-t}) \,dt \ge 0$ \citep[\S 6.2]{OlverEtAl10}. Thus,
    \[
        J_k = e^{-\gamma} e^{-T_k/\theta} e^{\Ein(J_k)} \ge e^{-\gamma} e^{-T_k/\theta},
    \]
    and (\ref{eq:tail}) follows from $\mathbb{E}\left(e^{-T_k/\theta}\right) = (\theta/(1+\theta))^{k}$. As regards the asymptotic statement, note that under the distribution of $T_k$ tilted by $e^{-T_k/\theta}$, i.e., under the gamma distribution with shape $k$ and rate $1 + 1/\theta$, we have $J_k \rightarrow 0$ in probability as $k \rightarrow \infty$. Moreover, the inequalities $\Ein(x) \le x$ and $E_{1}(x) < e^{-x} \log(1 + 1/x)$ \citep[\S 6.8]{OlverEtAl10} yield $e^{\Ein(J_k)} \le e + \theta/T_k$, which is uniformly integrable under the tilted distributions. Hence, the expectation of $e^{\Ein(J_k)}$ under the tilted distribution tends to one, as desired.
\end{proof}

\begin{remark} \label{rem:history}
    For $\theta = 1$, formula (\ref{eq:sparseProfile}) and the asymptotic form of (\ref{eq:tail}) were obtained by \citet{SheppLloyd66} for the cycle lengths of a uniformly random permutation. For general $\theta$, see \citet{Watterson76}, \citet{Griffiths88}, and \citet{MendoncaNegret26}.
\end{remark}

Note that $\ell_k(\theta)$ does not involve $n$, i.e., in the sparse regime the expected shares of the leading candidates stabilize as the list lengthens. Moreover, by (\ref{eq:tail}), the tail of the sparse profile is geometric with ratio $\theta/(1+\theta)$.

It remains to specify the distribution of $\theta$. We take it to be given by the measure $\theta^{-1} \,d\theta$, i.e., by the prior that is invariant under changes of scale and thus expresses ignorance of the order of magnitude of a positive quantity \citep{Jeffreys46,Jaynes68}. Since that measure is not finite, we first restrict it to an interval $[\theta_0, \Theta]$ and then let $\Theta \rightarrow \infty$. Formally, we assume that:

\begin{enumerate}[topsep=0pt,leftmargin=*,label=(\textbf{A\arabic*}),start=3]
    \item \label{A3} (random concentration) the total concentration $\theta$ is a random variable with density proportional to $\theta^{-1}$ on $[\theta_0, \Theta]$, where $0 < \theta_0 < \Theta < \infty$, and \textbf{A2} holds conditionally on $\theta$ with $\alpha = \theta/n$.
\end{enumerate}

\begin{definition}[Mixture weights] \label{def:mixture}
    Fix any $\theta_0 > 0$. The \emph{mixture weight} of the $k$-th position, where $k \in \mathbb{N}_{+}$, is
    \begin{equation}
        g_k(\theta_0) := \int_{\theta_0}^{\infty} \ell_k(\theta) \theta^{-1} \,d\theta.  \label{eq:mixtureDef}
    \end{equation}
\end{definition}

\begin{theorem}[Random concentration] \label{thm:random}
    Assume that \textbf{A1}--\textbf{A3} hold. Then:
    \begin{enumerate}
        \item[(i)] for every $k \in \mathbb{N}_{+}$, the expected share of the candidate in the $k$-th position relative to that of the leader satisfies
        \begin{equation}
            \lim_{\Theta \rightarrow \infty} \lim_{n \rightarrow \infty} \frac{\mathbb{E}(V_{(k)})}{\mathbb{E}(V_{(1)})} = \frac{g_k(\theta_0)}{g_1(\theta_0)};  \label{eq:relative}
        \end{equation}

        \item[(ii)] for every $k \ge 2$,
        \begin{equation}
            g_k(\theta_0) = \frac{1}{k-1} \sum_{j=1}^{k-1} \ell_j(\theta_0),  \label{eq:mixture}
        \end{equation}
        and in particular $\int_{0}^{\infty} \ell_k(\theta) \theta^{-1} \,d\theta = (k-1)^{-1}$;

        \item[(iii)] the mixture weight of the leader is finite and equals
        \begin{equation}
            g_1(\theta_0) = \int_{0}^{\infty} e^{-x} E_{1}\left(\theta_0 E_{1}(x)\right) \,dx.  \label{eq:leader}
        \end{equation}
    \end{enumerate}
\end{theorem}

\begin{proof}
    We begin with (ii) and (iii). Since all integrands are nonnegative, the order of integration in (\ref{eq:mixtureDef}) and (\ref{eq:sparseProfile}) can be exchanged. Fix any $k \ge 2$ and $y > 0$. Then $\int_{\theta_0}^{\infty} \theta^{k-2} e^{-\theta y} \,d\theta = \Gamma(k-1, \theta_0 y) / y^{k-1}$, and $\Gamma(k-1, \lambda)/(k-2)! = \sum_{j=0}^{k-2} e^{-\lambda} \lambda^{j}/j!$ for every $\lambda > 0$. Putting $y := E_{1}(x)$ and $\lambda := \theta_0 E_{1}(x)$, we see that the $j$-th term of the sum integrates against $e^{-x} \,dx$ to $\ell_{j+1}(\theta_0)$, which yields (\ref{eq:mixture}). Letting $\theta_0 \rightarrow 0$, we obtain $(k-1)^{-1}$, as desired. For $k = 1$ and every $y > 0$ we have $\int_{\theta_0}^{\infty} \theta^{-1} e^{-\theta y} \,d\theta = E_{1}(\theta_0 y)$, which yields (\ref{eq:leader}). The latter integral is finite. Indeed, recall that $\frac{1}{2} e^{-x} \log(1 + 2/x) < E_{1}(x) < e^{-x} \log(1 + 1/x)$ for every $x > 0$ \citep[\S 6.8]{OlverEtAl10}. Hence, $E_{1}\left(\theta_0 E_{1}(x)\right) < \log\left(1 + (\theta_0 E_{1}(x))^{-1}\right)$, and the right-hand side is bounded for $x \le 1$ and grows at most affinely in $x$ as $x \rightarrow \infty$.

    It remains to prove (i). Fix any $k \in \mathbb{N}_{+}$ and any $\Theta > \theta_0$. By \textbf{A3}, on a list of $n$ candidates
    \[
        \mathbb{E}(V_{(k)}) = \frac{1}{\log(\Theta/\theta_0)} \int_{\theta_0}^{\Theta} \pi^{(n)}_k(\theta/n) \theta^{-1} \,d\theta.
    \]
    Since $0 \le \pi^{(n)}_k(\theta/n) \le 1$, by Proposition~\ref{prop:sparse} and the bounded convergence theorem the right-hand side converges to $(g_k(\theta_0) - g_k(\Theta))/\log(\Theta/\theta_0)$ as $n \rightarrow \infty$. Therefore,
    \[
        \lim_{n \rightarrow \infty} \frac{\mathbb{E}(V_{(k)})}{\mathbb{E}(V_{(1)})} = \frac{g_k(\theta_0) - g_k(\Theta)}{g_1(\theta_0) - g_1(\Theta)}.
    \]
    Since $g_k(\theta_0)$ is finite for every $k \in \mathbb{N}_{+}$ by (ii) and (iii), we have $g_k(\Theta) \rightarrow 0$ as $\Theta \rightarrow \infty$, and (\ref{eq:relative}) follows.
\end{proof}

\begin{proposition}[Power priors] \label{prop:power}
    For every $s \in (0, 1)$ and every $k \in \mathbb{N}_{+}$,
    \begin{equation}
        \int_{0}^{\infty} \ell_k(\theta) \theta^{-s} \,d\theta = C_s \frac{\Gamma(k-s)}{\Gamma(k)}, \qquad \text{where} \qquad C_s := \int_{0}^{\infty} e^{-x} E_{1}(x)^{s-1} \,dx < \infty.  \label{eq:power}
    \end{equation}
\end{proposition}

\begin{proof}
    Fix any $s \in (0, 1)$, $k \in \mathbb{N}_{+}$, and $y > 0$. Then $\int_{0}^{\infty} \theta^{k-1-s} e^{-\theta y} \,d\theta = \Gamma(k-s)/y^{k-s}$. Putting $y := E_{1}(x)$ and exchanging the order of integration in (\ref{eq:sparseProfile}), we obtain (\ref{eq:power}). The constant $C_s$ is finite, because $E_{1}(x)^{s-1} e^{-x}$ behaves as $x^{1-s} e^{-sx}$ at infinity and as $(-\log x)^{s-1}$ at zero.
\end{proof}

\begin{remark} \label{rem:rate}
    By (\ref{eq:mixture}), for every $k \ge 2$ the mixture weight $g_k(\theta_0)$ is $(k-1)^{-1}$ times a partial sum of the series $\sum_{j=1}^{\infty} \ell_j(\theta_0) = 1$, the tail of which is geometric by Proposition~\ref{prop:sparse}. Hence, beyond the first few positions, $g_k(\theta_0) \approx (k-1)^{-1}$, whatever the lower bound, and the weights decay as $1/k$. Intuitively, averaging geometric profiles over a scale-invariant range of concentrations produces the harmonic rate of decay. This is an instance of the general mechanism by which a mixture of exponentially decaying terms with a scale-free distribution of rates decays polynomially \citep{Newman05}, and its elementary prototype is the identity $\int_{0}^{1} r^{k-1} \,dr = 1/k$. By Proposition~\ref{prop:power}, the exponent of the decay is inherited from the exponent of the prior, and the scale-invariant prior is the one that produces exponent one.
\end{remark}

\begin{remark} \label{rem:leader}
    Formula (\ref{eq:mixture}) does not extend to the leader, and no position receives an unbounded weight. Indeed, by Theorem~\ref{thm:random}(iii), the mixture weight of the leader is finite for every $\theta_0 > 0$, e.g., $g_1(\theta_0)$ equals $1.17$, $3.04$, and $5.28$ for $\theta_0 = 1$, $1/10$, and $1/100$, respectively. Since $E_{1}(z) = -\gamma - \log z + \Ein(z)$ and $0 \le \Ein(z) \le z$, the difference $g_1(\theta_0) - \log(1/\theta_0)$ remains bounded as $\theta_0 \rightarrow 0$, i.e., the weight of the leader grows as $\log(1/\theta_0)$. Intuitively, for small $\theta$ the leader receives nearly the whole vote, and the measure $\theta^{-1} \,d\theta$ assigns infinite mass to every neighborhood of zero. Accordingly, a positive lower bound is indispensable, and it is this bound that fixes the weight of the leader relative to the other positions. Through the partial sums in (\ref{eq:mixture}), it also affects the first few positions below the leader.
\end{remark}

A natural lower bound is $\theta_0 = 1$, which admits all lists with a limiting effective number of candidates of at least two. Then $g_1(1) = 1.171$, and the weights of the second to fifth positions relative to that of the leader are $0.533$, $0.356$, $0.263$, and $0.206$, against $1/2$, $1/3$, $1/4$, and $1/5$ for the harmonic weights (Table~\ref{tbl:relative}; Figure~\ref{fig:profiles}, green curve). Note that the product $k g_k(1)$, which would be constant for a harmonic profile, equals $1.17$, $1.25$, $1.25$, $1.23$, and $1.20$ for $k = 1, \dots, 5$ and tends to one as $k \rightarrow \infty$. Indeed, at the first positions below the leader the factor $k/(k-1)$ is offset by the partial sum in (\ref{eq:mixture}), which is smaller than one. Thus, the weights implied by random concentration with this lower bound are closely approximated by the harmonic weights, which have the advantage of a closed form. Renormalized over the top $m$ positions, the two sets of weights are at a distance of $0.014$ to $0.034$ for every $m$ between 2 and 100 (Table~\ref{tbl:topM}). The agreement depends on the lower bound: the distance is minimized for $\theta_0$ between $0.8$ and $1.2$, and it rises to between $0.02$ and $0.09$ for $\theta_0 = 1/2$ and to between $0.04$ and $0.06$ for $\theta_0 = 2$.

\begin{table}[!htb]
    \centering\small
    \caption{Expected share of the $k$-th candidate relative to the leader.}
    \label{tbl:relative}
    \begin{tabular}{lcccccccc}
    \toprule
    & $k = 1$ & $2$ & $3$ & $4$ & $5$ & $6$ & $7$ & $8$ \\
    \midrule
    Harmonic, $1/k$ & 1.000 & 0.500 & 0.333 & 0.250 & 0.200 & 0.167 & 0.143 & 0.125 \\
    Uniform, $\alpha = 1$, $n = 30$ & 1.000 & 0.750 & 0.625 & 0.541 & 0.479 & 0.428 & 0.387 & 0.351 \\
    Sparse, $\theta = 3$ & 1.000 & 0.507 & 0.313 & 0.208 & 0.144 & 0.102 & 0.073 & 0.053 \\
    $\theta$ log-uniform on $[1, \infty)$ & 1.000 & 0.533 & 0.356 & 0.263 & 0.206 & 0.168 & 0.141 & 0.121 \\
    \bottomrule
    \end{tabular}
\end{table}

\begin{table}[!htb]
    \centering\small
    \caption{Total variation distance $d$ between the harmonic weights and the mixture weights, both renormalized over the top $m$ positions. The parameter $\theta_0^*$ minimizes the distance.}
    \label{tbl:topM}
    \begin{tabular}{rccc}
    \toprule
    $m$ & $d$ at $\theta_0 = 1$ & $\theta_0^*$ & $d$ at $\theta_0^*$ \\
    \midrule
    3 & 0.016 & 0.77 & 0.003 \\
    5 & 0.014 & 0.84 & 0.007 \\
    10 & 0.016 & 0.92 & 0.014 \\
    20 & 0.020 & 1.00 & 0.020 \\
    30 & 0.025 & 1.03 & 0.023 \\
    50 & 0.030 & 1.07 & 0.025 \\
    100 & 0.034 & 1.16 & 0.027 \\
    \bottomrule
    \end{tabular}
\end{table}

\begin{remark} \label{rem:approx}
    The agreement between the mixture weights and the harmonic weights is approximate. Below the leader the former are close to $(k-1)^{-1}$ rather than $1/k$, so that, relative to the leader, the mixture places up to 7 percent more weight than the harmonic profile on the second to fifth positions and up to 15 percent less on distant ones.
\end{remark}

\begin{remark} \label{rem:finite}
    Theorem~\ref{thm:random} is obtained in the limit in which the list is the leading segment of an unbounded pool of candidates, so that its lowest positions are not depressed by the end of the list. On a list of exactly $n$ candidates with $\alpha$ log-uniform on $[1/n, A]$, the distance from the harmonic weights depends on the upper bound. For $n$ between 5 and 50 it is $0.16$ to $0.19$ if $A = 1$, in which case the lowest positions receive far less than their harmonic weight, and $0.035$ to $0.058$ if $A$ is between 8 and 32, i.e., if the prior extends to divisions of the vote much more even than those typical of the uniform distribution. Thus, the approximation rests on the premise, shared with the harmonic scheme itself, that the ratios between the weights of positions do not depend on the length of the list.
\end{remark}

\begin{table}[!htb]
    \centering\small
    \caption{Total variation distance $d$ between the harmonic weights and the expected-share profile of a symmetric Dirichlet distribution on a list of $n$ candidates. The parameter $\alpha^*$ minimizes the distance.}
    \label{tbl:single}
    \begin{tabular}{rccccc}
    \toprule
    $n$ & $d$ at $\alpha = 1/2$ & $d$ at $\alpha = 1$ & $\alpha^*$ & $n\alpha^*$ & $d$ at $\alpha^*$ \\
    \midrule
    3 & 0.155 & 0.071 & 1.81 & 5.4 & 0.028 \\
    5 & 0.149 & 0.067 & 1.17 & 5.8 & 0.053 \\
    10 & 0.126 & 0.099 & 0.66 & 6.6 & 0.092 \\
    20 & 0.134 & 0.146 & 0.59 & 11.8 & 0.134 \\
    30 & 0.153 & 0.175 & 0.54 & 16.2 & 0.152 \\
    50 & 0.185 & 0.216 & 0.34 & 17.2 & 0.178 \\
    100 & 0.232 & 0.269 & 0.19 & 19.4 & 0.217 \\
    \bottomrule
    \end{tabular}
\end{table}

No fixed value of the parameter leads to similar weights. With a fixed total concentration, the sparse profile can agree with the harmonic weights only at the top of the list, since its tail is geometric (Table~\ref{tbl:relative}, third row; Figure~\ref{fig:profiles}, orange curve). For instance, for $\theta = 3$ the two differ by $d = 0.010$ over the top three positions, but even for the best-fitting value of $\theta$ the distance grows from $0.007$ for three positions to $0.173$ for one hundred. With a fixed $\alpha$ on a list of $n$ candidates the agreement is no better (Table~\ref{tbl:single}). For $n = 2$ the single free parameter suffices, as $\pi^{(2)}_1(\alpha) = 2/3$ for $\alpha \approx 2.60$. For longer lists it does not. The best fit is obtained for $\alpha > 1$ on very short lists and for $\alpha < 1$ on longer ones; it deteriorates steadily with list length, and neither the best-fitting parameter nor the corresponding total concentration is stable across list lengths. The cause is the one identified in Section~\ref{secFixed}: on doubly logarithmic axes, a profile whose elasticity rises from zero to infinity can only touch a harmonic line, and it is flatter than harmonic above the point of tangency and steeper below it.

\section{\label{secDiscussion}Discussion}

We have derived a priori weights of candidates on ordered lists from a probabilistic model of the division of the list vote. The uniform distribution on the simplex -- the natural a priori model of that division -- divides the vote too evenly, and under it the leading candidates become more alike as the list lengthens. The other members of the symmetric Dirichlet family require a concentration parameter, and any fixed value of that parameter is a substantive assumption about lists that can be expected to differ across countries, parties, and elections. If the parameter is treated as random instead, with the prior appropriate to a quantity of unknown order of magnitude, the expected shares decay as $1/k$ (Theorem~\ref{thm:random}). Moreover, if the prior admits all lists with at least two effective candidates, the harmonic weights reproduce those shares to within a few percent of the total weight. Accordingly, the harmonic weights are the a priori weights appropriate when neither the allocation of votes among candidates nor its degree of concentration is known.

Note that the harmonic scheme is connected with the Dirichlet family in two further ways. It is exactly the profile of normalized expected margins under the uniform distribution (Proposition~\ref{prop:margins}), and for the leading three positions it approximates the sparse profile with an effective number of candidates close to four. Nevertheless, it is not the expected-share profile of any symmetric Dirichlet distribution with a fixed parameter, however concentrated.

The above conclusions are subject to several limitations. First, they depend on the prior in two ways. By Proposition~\ref{prop:power}, a prior proportional to $\theta^{-s}$ leads to weights that decay as $k^{-s}$, so that the harmonic rate of decay is specific to the scale-invariant case, and the weight of the leader relative to the other positions is fixed by the lower bound of the prior (Remark~\ref{rem:leader}). Second, the analysis rests on Assumption \textbf{A1}, whereas actual lists are sorted imperfectly, i.e., candidates placed lower sometimes outpoll those placed above them. Under imperfect sorting the expected share of a list position is a mixture of the expected shares of several ranks, wherefore the profile by position is flatter than the profile by rank. Third, Theorem~\ref{thm:random} concerns the leading positions of a long list, and on a list considered in isolation the approximation requires a prior that extends to nearly even divisions of the vote (Remark~\ref{rem:finite}). Fourth, the symmetric Dirichlet family is only one family of exchangeable models. Polynomial decay of ranked shares also arises in the two-parameter Poisson--Dirichlet family \citep{PitmanYor97}, in which, however, the exponent exceeds one, so that the harmonic rate is attained only at the boundary.

Our analysis is theoretical, i.e., the weights are derived a priori rather than estimated from electoral returns, and it differs in three respects from the best-documented empirical account of how votes are divided within open lists -- the scaling relation of \citet{FortunatoCastellano07}. For elections in Italy, Poland, and Finland they find that the distribution of a candidate's votes relative to the average of his or her list is the same across lists and elections and is well described by a lognormal distribution, and \citet{ChatterjeeEtAl13} confirm the regularity for countries with similar open-list rules. The first and principal difference concerns the status of the distribution. Their lognormal distribution is fitted to results rather than derived. Moreover, the word-of-mouth model by which they account for the regularity is studied by simulation only, and it is not shown to yield a lognormal distribution. Indeed, an analytic treatment of an approximation of that model leads to a power-law upper tail on long lists, and such a tail, rather than a lognormal one, is reported for lists of more than thirty candidates in Italy, although not in Finland \citep{PalombiToti16}. By contrast, the profiles derived here follow from explicitly stated assumptions and are available before any votes are cast. The second difference concerns the object. Their relation describes the performance of a single candidate, pooled over all candidates of all lists, whereas the weights sought here are expectations of ordered shares, which depend on the joint distribution within a list and follow from their relation only under a further assumption, such as the independence of the candidates' strengths. The third difference concerns list length. Under their relation the distribution of a candidate's share relative to the list average does not depend on the length of the list. This corresponds to a fixed per-candidate parameter, as in Section~\ref{secFixed}, with a lognormal distribution in place of the gamma distribution. The sparse regime, on which Theorem~\ref{thm:random} is built, holds the total concentration fixed instead, and under it that distribution becomes more skewed as lists lengthen. Hence, the two models cannot both hold exactly.

The practical consequences of the last difference depend on the length of the list. If the strengths of the candidates are taken to be independent and lognormal with the fitted log-variance of $1.08$, the leader's expected share is $0.33$, $0.18$, and $0.09$ on lists of $10$, $30$, and $100$ candidates, respectively, against harmonic weights of $0.34$, $0.25$, and $0.19$ and expected shares of $0.29$, $0.13$, and $0.05$ under the uniform distribution. The total variation distance of the harmonic weights from the implied profile grows from $0.05$ to $0.19$ over this range, while that of the uniform profile stays below $0.08$. Thus, harmonic weights agree with the empirical regularity to within about 5 percent of the total weight for lists of up to ten candidates, but not for long lists, on which the regularity implies a faster decline of the shares of the leaders than the harmonic weights allow. Note, however, that the relation was established on pooled distributions, without separate examination of the dependence on list length, and that departures from it have been noted in the upper tail, in which the leaders are found \citep{ChatterjeeEtAl13}. In particular, as noted above, the lognormal form has been found to fail on long lists, i.e., precisely where the above comparison is least favorable to the harmonic weights.

\section*{\label{secNumerics}Appendix. Numerical Methods}

The expected order statistics in (\ref{eq:levels}) were computed as
\[
    \mathbb{E}(G_{(k)}) = \int_{0}^{\infty} Q_\alpha(1 - e^{-s}) \varphi_{k:n}(s) \,ds, \qquad \text{where} \qquad \varphi_{k:n}(s) := \frac{n!}{(k-1)!\,(n-k)!} (1 - e^{-s})^{n-k} e^{-ks}
\]
is the density of the $k$-th largest of $n$ unit exponential variables, by Simpson's rule on 60{,}001 nodes refined near zero. The results reproduce (\ref{eq:uniform}) to within $10^{-14}$, sum to one over $k$ to within $10^{-13}$, satisfy the second equality in (\ref{eq:spacings}) when its right-hand side is evaluated independently, and agree with Monte Carlo estimates ($4 \times 10^{5}$ draws) to within $2 \times 10^{-4}$. Formula (\ref{eq:sparseProfile}) was integrated on a logarithmic grid; the Golomb--Dickman constant $\ell_1(1) = 0.62433\dots$ is recovered to five decimal places, and the values agree to four decimal places with the Dirichlet profile for $n = 5000$ and $\alpha = \theta/n$. The identities of Theorem~\ref{thm:random} were checked by direct numerical integration over $\theta$ and by simulation of the hierarchical model ($\theta$ log-uniform on $[1, 50]$, $n = 3000$, $12{,}000$ draws), for which the simulated expected shares of the top five candidates, $(0.277, 0.143, 0.093, 0.066, 0.050)$, agree to within simulation error with the values $(0.279, 0.144, 0.093, 0.066, 0.050)$ implied by (\ref{eq:mixture}) and (\ref{eq:leader}). Minimizing parameters were found by a grid search followed by bounded scalar minimization. The profiles for a random parameter on a finite list were obtained by Gauss--Legendre quadrature with 48 nodes in $\log \alpha$. The lognormal comparison of Section~\ref{secDiscussion} was obtained by simulation, with $10^{6}$ draws for $n \le 30$ and $5 \times 10^{5}$ draws for longer lists; the standard errors of the expected shares do not exceed $1.5 \times 10^{-4}$. All computations were carried out in Python with SciPy \citep{VirtanenEtAl20}; the code is provided as an ancillary file.

\bibliographystyle{plainnat}
\bibliography{refs}

\end{document}